# EFFECTIVE MEDIUM THEORY FOR THE THERMOELECTRIC PROPERTIES OF COMPOSITE MATERIALS WITH VARIOUS PERCOLATION THRESHOLDS

**Andrei A. Snarskii**[1, 2, *], **Pavel Yuskevich**[1]

[1] National Technical University of Ukraine "Igor Sikorsky Kyiv Polytechnic Institute", Prospekt Peremohy 37, 03056 Kiev, Ukraine; asnarskii@gmail.com (A.S); pasha.yusk@gmail.com (P.Y.)
[2] Institute for Information Recording, NAS of Ukraine, Mykoly Shpaka Street 2, 03113 Kiev, Ukraine

* Correspondence: asnarskii@gmail.com (A.S)

*In the work, a modified effective medium theory is constructed for calculating the effective properties of thermoelectric composites with different values of percolation thresholds. It is shown that even at concentrations beyond the critical region, the threshold value is essential for determining the effective properties. Two fundamentally different cases of a set of local properties of the composite are considered. In one of these cases, the conductivity and thermal conductivity of one of the phases is simultaneously greater than the conductivity and thermal conductivity of the other phase. The second, anomalous case, when the electrical conductivity of the first phase ($\sigma_1$) is greater than that of the second, but the thermal conductivity of the first phase is less than that of the second, shows unusual concentration behavior of effective conductivity, i.e. with an increase in the well-conducting phase, the effective conductivity $\sigma_e$ shows a decrease (rather than growth as in the standard case, see Fig. 1a), which at $p \approx \tilde{p}_c$ goes over to growth. Bibl. 5, Fig. 5.*

## Introduction

The calculation of effective values for composite materials is a complex problem that cannot be solved in the general case. Solutions are also possible as an exceptional case, for one-dimensional inhomogeneity, or for strictly periodic structures, for example, for spherical inclusions of one phase in the matrix of the other. Even in the case of simple-shaped inclusions, the solutions are rather bulky expressions of infinite series that are difficult to analyze [1 – 7].

To describe randomly inhomogeneous environment with randomly located inclusions of one phase in the other, there are approximate methods that allow one to approximately describe the concentration behavior of effective coefficients with different accuracy. For example, the Maxwell approximation [8] allows describing the concentration behavior of effective coefficients accurate to the first degree of concentration.

For the entire range of concentrations, a good approximation is the Bruggeman – Landauer approximation [9, 10], which is a self-consistency method (effective medium approximation).

The lack of the Bruggeman-Landauer approximation is the percolation threshold fixed in this approximation. With a large difference in the physical properties of the phases, for example, when the conductivity of the first phase $\sigma_1$ is much larger than that of the second $\sigma_2 \cdot (\sigma_1/\sigma_2) \gg 1$, a sharp change in the behavior of the effective conductivity $\sigma_e$ with a change in the concentration of phases occurs when the concentration of the first phase $\sigma_1$ is equal to $p_c=1/3$. At the same time, the value of the percolation threshold $p_c$ of real composites can be different depending on their structure, which is related to their fabrication technique.

In [11], a modification of the Bruggeman – Landauer approximation was presented, which allows one to specify any percolation threshold $\tilde{p}_c$ for the problem of calculating the effective conductivity.

The objective of this work is to modify the self-consistent approximation for calculating the effective thermoelectric properties of composites (Bruggeman-Landauer approximation for thermoelectric phenomena) and to show the influence of the assigned percolation threshold on the effective properties of two-phase thermoelectric composites.

The article is organized as follows: in the first section, we consider the Bruggeman-Landauer approximation and its modification based on [11]. In the second section, the system of thermoelectric equations is rewritten in a convenient form for constructing the Bruggeman-Landauer approximation and a modification of the approximation is proposed. From the obtained results it follows that the thermoelectric figure of merit, depending on the percolation threshold, has a maximum, which is an interesting consequence for experimental verification. In the third section, we present the calculation of effective properties in the "anomalous" case, where an unusual behavior of the effective conductivity is observed.

## Modification of the Bruggeman-Landauer approximation in the effective conductivity problem

The Bruggeman-Landauer approximation can be written as

$$\frac{\sigma_e - \sigma_1}{2\sigma_e + \sigma_1} p + \frac{\sigma_e - \sigma_2}{2\sigma_e + \sigma_2}(1-p) = 0, \qquad (1)$$

where $\sigma_1$ is the first phase conductivity, $\sigma_2$ is the second phase conductivity.

The effective conductivity obtained from solution (1) describes well the entire concentration range. With greater inhomogeneity, at $\sigma_1/\sigma_2 \gg 1$ close to concentration $p=p_c=1/3$ there is a sharp change in the concentration behavior of $\sigma_e$, which qualitatively describes the percolation behaviour (an analogue of the second-order phase transition). Naturally, such a "simple" approximation as (1) cannot completely describe percolation laws. For example, the critical indices $\sigma_e$ close to $p_c$ obtained from (1) [1, 11]. Their numerical value $t=q=1$ differs from the percolation ($t=2$ and $q=0.73$).

According to modification [11], the Bruggeman-Landauer approximation (1) is replaced by

$$\frac{\frac{\sigma_e - \sigma_1}{2\sigma_e + \sigma_1}}{1 + c(p, \tilde{p}_c)\frac{\sigma_e - \sigma_1}{2\sigma_e + \sigma_1}} p + \frac{\frac{\sigma_e - \sigma_2}{2\sigma_e + \sigma_2}}{1 + c(p, \tilde{p}_c)\frac{\sigma_e - \sigma_2}{2\sigma_e + \sigma_2}}(1-p) = 0, \qquad (2)$$

where $c(p, \tilde{p}_c)$, the Sarychev-Vinogradov term, is of the form

$$c(p, \tilde{p}_c) = (1 - 3\tilde{p}_c)\left(\frac{p}{\tilde{p}_c}\right)^{\tilde{p}_c}\left(\frac{1-p}{1-\tilde{p}_c}\right)^{1-\tilde{p}_c}, \qquad (3)$$

and $\tilde{p}_c$ is the preassigned percolation threshold.

According to (2), $\sigma_e(p)$ at $h=\sigma_1/\sigma_2 \to 0$ close to $\tilde{p}_c$ has the same exponential behavior, as in the standard Bruggeman-Landauer approximation (1)

$$\sigma_e(p) \sim \sigma_1\left(p - \tilde{p}_c\right)^t, \ \sigma_2 = 0, \ p > p_c,$$

$$\sigma_e(p) \sim \sigma_2\left(\tilde{p}_c - p\right)^{-q}, \ \sigma_1 = \infty, \ p < p_c, \qquad (4)$$

where critical indices $t=1$ and $q=1$.

## Modification of the Bruggeman-Landauer approximation for thermoelectric phenomena

In the case of thermoelectric phenomena we write the local relation between the electric current $j$, heat flux density $q$, temperature gradient $\nabla T$ and electric field strength $E$ in the form [8, 12]

$$\mathbf{j} = \sigma \mathbf{E} + \sigma \alpha (-\nabla T),$$

$$\frac{\mathbf{q}}{T} = \sigma \alpha \mathbf{E} + \kappa \frac{1+ZT}{T}(-\nabla T), \quad (5)$$

where $\sigma, \kappa$ are electric conductivity and thermal conductivity, $\alpha$ is differential thermoEMF, and

$$ZT = \frac{\sigma \alpha^2}{\kappa} T, \quad (6)$$

thermoelectric figure of merit (the Ioffe number) $Z$ multiplied by temperature.

Local kinetic coefficients $\sigma$, $\kappa$, $\alpha$ are coordinate-dependent and in the case of a two-phase composite take on the values $\sigma_1$, $\kappa_1$, $\alpha_1$ in the first phase and $\sigma_2$, $\kappa_2$, $\alpha_2$ - in the second.

The properties of composite as a whole are assigned by the effective kinetic coefficients relating by definition the volume average "currents" – electric $\mathbf{j}$ and thermal $\mathbf{q}$ to "forces" – electric field $\mathbf{E}$ and temperature gradient $\nabla T$

$$\langle \mathbf{j} \rangle = \sigma_e \langle \mathbf{E} \rangle + \sigma_e \alpha_e \langle -\nabla \mathrm{T} \rangle,$$

$$\frac{\langle \mathbf{q} \rangle}{T} = \sigma_e \alpha_e \langle \mathbf{E} \rangle + \kappa_e \frac{1+Z_e T}{T} \langle -\nabla \mathrm{T} \rangle, \quad (7)$$

where

$$Z_e = \frac{\sigma_e \alpha_e^2}{\kappa_e}.$$

Systems (5) and (7) can be written in matrix form, convenient for further consideration

$$\begin{pmatrix} \mathbf{j} \\ \langle \mathbf{q} \rangle / T \end{pmatrix} = \begin{pmatrix} \sigma & \sigma\alpha \\ \sigma\alpha & \kappa \frac{1+ZT}{T} \end{pmatrix} \begin{pmatrix} \langle \mathbf{E} \rangle \\ \langle -\nabla \mathrm{T} \rangle \end{pmatrix}. \quad (8)$$

Note that in this record, as it must be according to Onsager's principle [12, 13], the matrix of kinetic coefficients is symmetric. Similarly, for the effective values

$$\begin{pmatrix} \mathbf{j} \\ \frac{\langle \mathbf{q} \rangle}{T} \end{pmatrix} = \begin{pmatrix} \sigma_e & \sigma_e \alpha_e \\ \sigma_e \alpha_e & \kappa_e \frac{1+Z_e T}{T} \end{pmatrix} \begin{pmatrix} \langle \mathbf{E} \rangle \\ \langle -\nabla \mathrm{T} \rangle \end{pmatrix}. \quad (9)$$

Introduce a generalized current **i** and a generalized force **e**

$$\mathbf{i} = \begin{pmatrix} \mathbf{j} \\ \langle \mathbf{q} \rangle / T \end{pmatrix}, \quad \mathbf{e} = \begin{pmatrix} \langle \mathbf{E} \rangle \\ \langle -\nabla T \rangle \end{pmatrix}, \tag{10}$$

which are interrelated by the matrix of local kinetic coefficients $\hat{\Omega}$

$$\mathbf{i} = \hat{\Omega}\mathbf{e}, \quad \hat{\Omega} = \begin{pmatrix} \sigma & \sigma\alpha \\ \sigma\alpha & \kappa\dfrac{1+ZT}{T} \end{pmatrix}, \tag{11}$$

and similarly for volume average currents and forces

$$\langle \mathbf{i} \rangle = \hat{\Omega}_e \langle \mathbf{e} \rangle. \tag{12}$$

In matrix notation, a self-consistent approximation of thermoelectric problem (analogue of the Bruggeman-Landauer equation) can be written as [2]

$$\frac{\hat{\Omega}_e - \hat{\Omega}_1}{2\hat{\Omega}_e + \hat{\Omega}_1} p + \frac{\hat{\Omega}_e - \hat{\Omega}_2}{2\hat{\Omega}_e + \hat{\Omega}_2}(1-p) = 0, \tag{13}$$

where expressions of the type $1/\left(2\hat{\Omega}_e + \hat{\Omega}_1\right)$ are understood as multiplying on the right by the inverse matrix.

Here, we (similarly to [11]) propose the following modification of a self-consistent approximation (13) for thermoelectric problem

$$\frac{\dfrac{\hat{\Omega}_e - \hat{\Omega}_1}{2\hat{\Omega}_e + \hat{\Omega}_1}}{1 + c(p, \tilde{p}_c)\dfrac{\hat{\Omega}_e - \hat{\Omega}_1}{2\hat{\Omega}_e + \hat{\Omega}_1}} p + \frac{\dfrac{\hat{\Omega}_e - \hat{\Omega}_2}{2\hat{\Omega}_e + \hat{\Omega}_2}}{1 + c(p, \tilde{p}_c)\dfrac{\hat{\Omega}_e - \hat{\Omega}_2}{2\hat{\Omega}_e + \hat{\Omega}_2}}(1-p) = 0. \tag{14}$$

Note at once that at $\tilde{p}_c = 1/3$ Eq.(14) goes over to standard approximation (13).

Fig.1 shows the concentration dependences of $\sigma_e$, $\kappa_e$, $\alpha_e$, $Z_eT$ for various $\tilde{p}_c$.

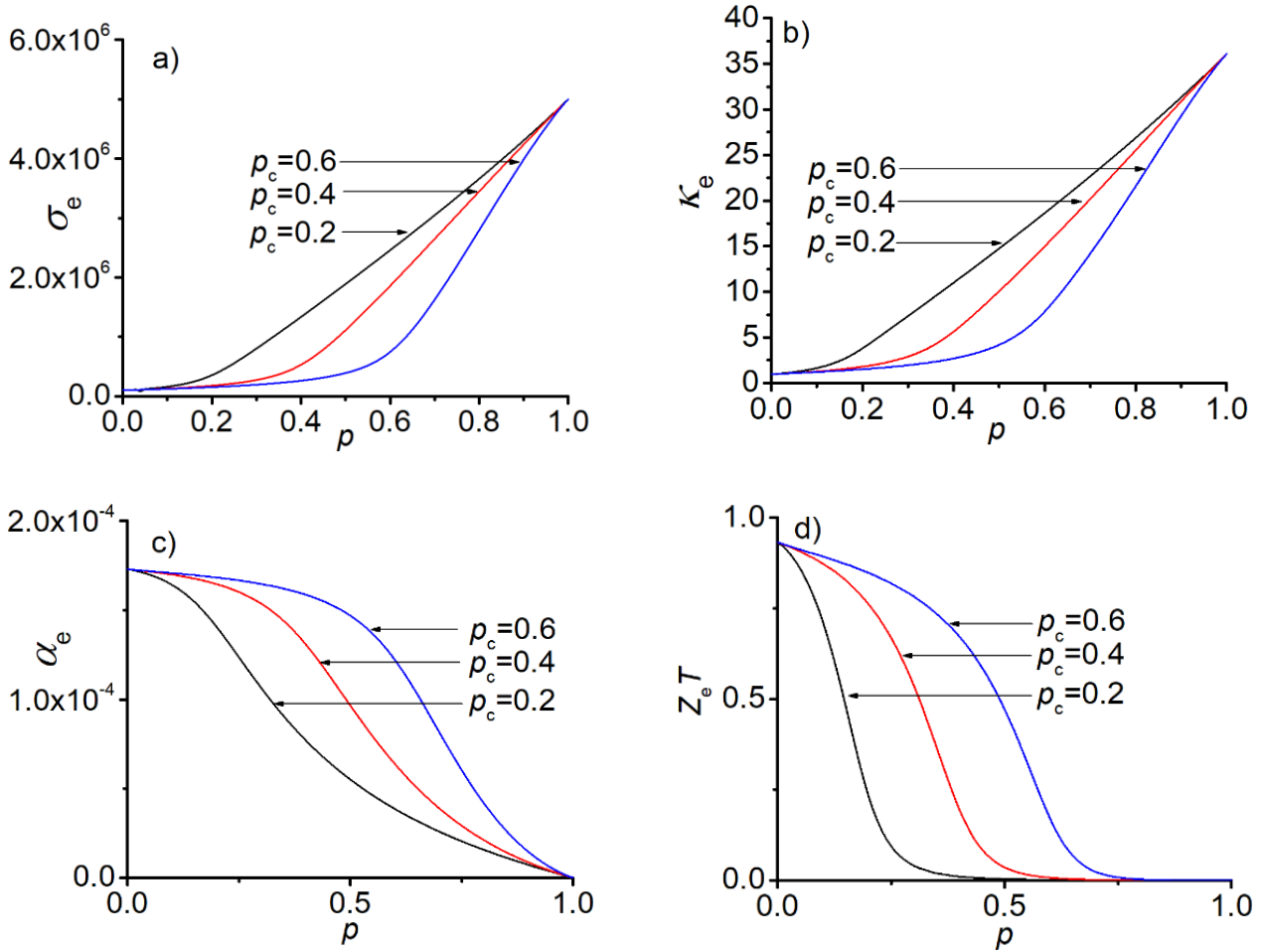

*Fig.1 Concentration dependences of effective conductivity, thermal conductivity, thermoEMF and figure of merit. The values of coefficients in the phases were selected [15]:*
*for the first phase – $\sigma_1 = 5 \cdot 10^6$ Ohm$^{-1}$m$^{-1}$, $\kappa_1 = 36.1$ W/(m·K), $\alpha_1 = 0$ V/K*
*for the second phase – $\sigma_2 = 10^5$ Ohm$^{-1}$m$^{-1}$, $\kappa_2 = 0.963$ W/(m·K), $\alpha_2 = 173 \cdot 10^6$*

As can be seen from Fig. 1, the effective conductivity increases with the addition of a better conductive phase, as it must be. Effective thermal conductivity behaves similarly. Accordingly, the effective thermoEMF decreases when a phase with a lower thermoEMF is added. It can also be seen from Fig. 1 that, for example, for effective conductivity, all lines behave identically, however, the region is shifted in which there is a sharp increase in conductivity, or, in other words, the percolation threshold is shifted. The thermoelectric figure of merit (the Ioffe number) decreases monotonically, as it must be, when the first phase with a lower thermoelectric figure of merit is added.

It can be strictly shown that in the case of $h_\sigma = \sigma_1/\sigma_2 \to 0$ and $h_\kappa = \kappa_1/\kappa_1 \to 0$, $\sigma_e$ and $\kappa_e$ either become equal to zero at $\sigma_2 = 0$ and $\kappa_2 = 0$ and $p \to \tilde{p}_c$, or diverge at $\sigma_1 = \infty$ and $\kappa_1 = \infty$ and $p \to \tilde{p}_c$.

As follows from (14), the values of effective coefficients at a given concentration $p$ of the first phase depend on the percolation threshold of composite $\tilde{p}_c$. The closer concentration $p$ to $\tilde{p}_c$, the more significant is this dependence, but even at lower concentrations, the difference occurs.

Fig. 2 shows the dependences of effective coefficients $\sigma_e$, $\kappa_e$, $\alpha_e$, $Z_eT$ on the value of $\tilde{p}_c$ for different values of $p$-concentration of the first phase.

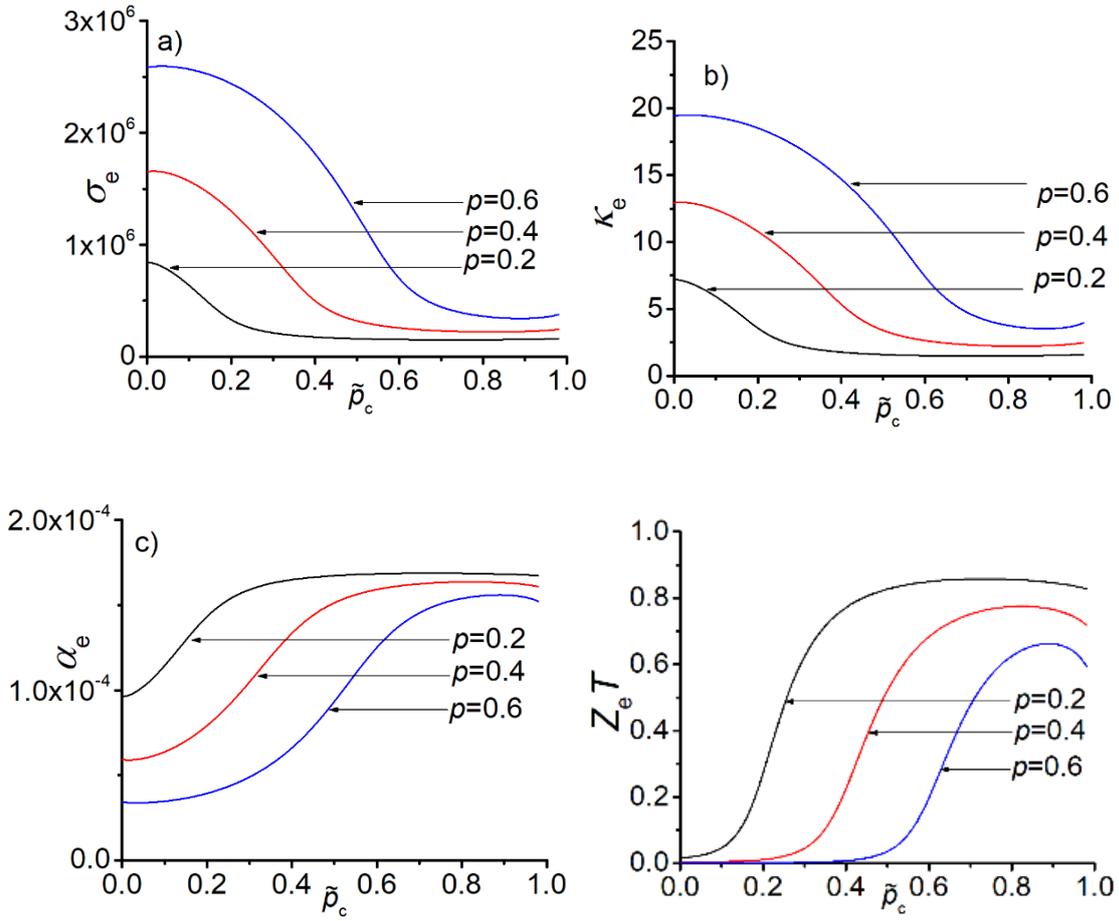

*Fig. 2 Dependences of effective kinetic coefficients on the percolation threshold $\tilde{p}_c$ at a given concentration p of the first phase. The values of coefficients in the phases were selected the same as in Fig. 1.*

Fig. 2 shows that the value of effective conductivity, at a given concentration, decreases with increasing the threshold $\tilde{p}_c$. The effective thermal conductivity behaves similarly. It is also noteworthy that, at a given concentration, with increasing the threshold $\tilde{p}_c$, the thermoelectric figure of merit (the Ioffe number) also increases, but at large $\tilde{p}_c$ the maximum is noticeable and the growth changes to a small drop. It would be interesting to verify the existence of such a maximum experimentally and to determine whether this maximum is an "artifact" of our modification.

**Modification of the Bruggeman-Landauer approximation for thermoelectric phenomena in the "anomalous" case**

Correction (3) introduced in the Bruggeman-Landauer approximation (2) suggests that the conductivity of the first phase is greater than the conductivity of the second phase $\sigma_1 > \sigma_2$. In this case, at $\tilde{p}_c$ given in (3), the percolation threshold that can be found when solving (2) will be equal to the preassigned $p_c = \tilde{p}_c$. Subject to the inverse inequality $\sigma_2 > \sigma_1$, correction (3) should be changed.

At $\alpha_1 = \alpha_{e2} = 0$, i.e. in the absence of thermoelectric phenomena, the effective medium approximations (13) or (14) are divided into two independent – one for electric conductivity, the other – for thermal conductivity. In the case when σ$_1$ > σ$_2$ and κ$_1$ > κ$_2$ these independent equations includes the same correction $c(p, \tilde{p}_c)$. However, in the opposite case, when σ$_1$ > σ$_2$, but κ$_1$ > κ$_2$, corrections for σ$_e$ and κ$_e$ should be different. In the general case, when α$_1$ ≠ α$_2$, this difference remains and modification (13) (14) assumes a more complex form.

$$\frac{\dfrac{\hat{\Omega}_e - \hat{\Omega}_1}{2\hat{\Omega}_e + \hat{\Omega}_1}}{1 + \hat{C}(p,\tilde{p}_c)\dfrac{\hat{\Omega}_e - \hat{\Omega}_1}{2\hat{\Omega}_e + \hat{\Omega}_1}} p + \frac{\dfrac{\hat{\Omega}_e - \hat{\Omega}_2}{2\hat{\Omega}_e + \hat{\Omega}_2}}{1 + \hat{C}(p,\tilde{p}_c)\dfrac{\hat{\Omega}_e - \hat{\Omega}_2}{2\hat{\Omega}_e + \hat{\Omega}_2}} (1-p) = 0, \qquad (15)$$

where

$$\hat{C}(p,\tilde{p}_c) = \begin{pmatrix} c_\sigma(p,\tilde{p}_c) & 0 \\ 0 & c_\kappa(p,\tilde{p}_c) \end{pmatrix}, \qquad (16)$$

and corrections $c_\sigma(p,\tilde{p}_c)$ and $c_\kappa(p,\tilde{p}_c)$ in (16), depending on the ratios σ$_1$ / σ$_2$ и κ$_1$ / κ$_2$ are of different form. In the case when σ$_1$ > σ$_2$ and κ$_1$ < κ$_2$ the correction $c_\sigma(p,\tilde{p}_c)$ remains the same – (3), and $c_\kappa(p,\tilde{p}_c)$ is given by

$$c_\kappa(p,\tilde{p}_c) = [1 - 3(1-\tilde{p}_c)] \left(\frac{p}{\tilde{p}_c}\right)^{\tilde{p}_c} \left(\frac{1-p}{1-\tilde{p}_c}\right)^{1-\tilde{p}_c}. \qquad (17)$$

Fig. 3 shows the concentration dependences of σ$_e$, κ$_e$, α$_e$, $Z_eT$ for different values of percolation threshold.

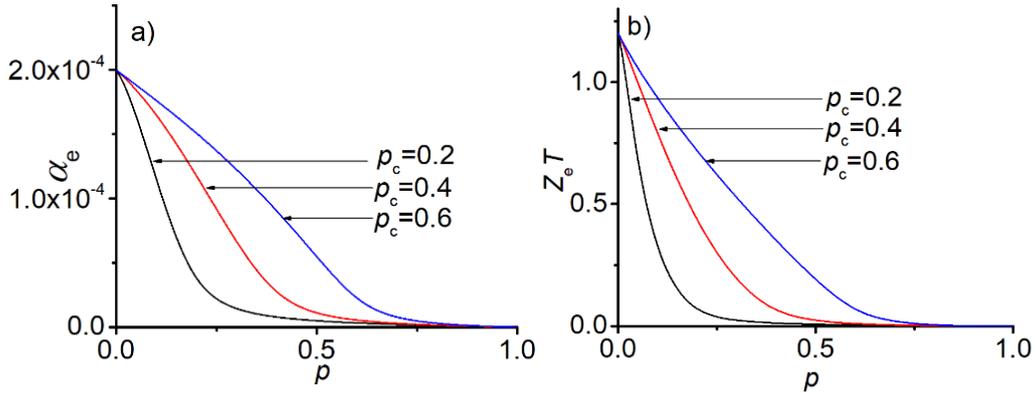

*Fig. 3. Dependences of effective thermoEMF and figure of merit on the concentration p of the first phase at a given percolation threshold $\tilde{p}_c$. The values of coefficients in the phases were selected the same as in Fig. 1.*

One of such materials for which the Wiedemann-Franz law – high conductivity, but low thermal conductivity - is violated significantly, is described in [16].
Fig.4 shows the dependences of effective properties on the value of percolation threshold.

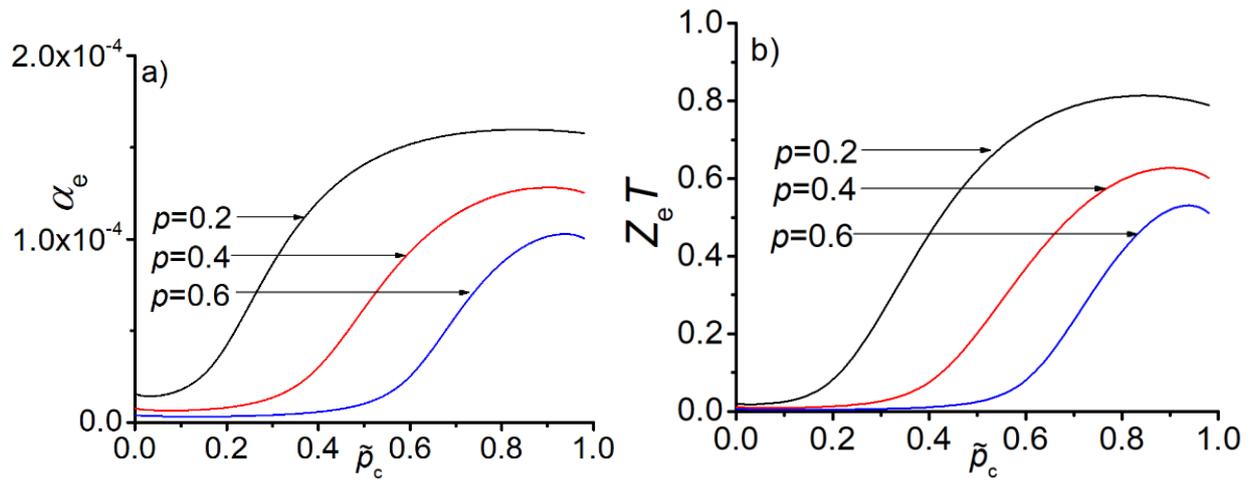

*Fig. 4. Dependences of effective thermoEMF and figure of merit on the percolation threshold $\tilde{p}_c$ at a given concentration $p$ of the first phase. The values of coefficients in the phases were elected the same as in Fig. 1.*

Just as in the usual case, the effective figure of merit has a maximum.

**Discussion**

    The figure below shows the dependences of effective conductivity and thermal conductivity on the concentration and percolation threshold in the "anomalous" case. It should be noted that effective conductivity has an unusual behavior: when a phase with good conductivity is added, effective conductivity first decreases, and then begins to grow. It seems interesting to experimentally verify whether this is a defect in the theory, or whether such conductivity is actually observed in real composites.

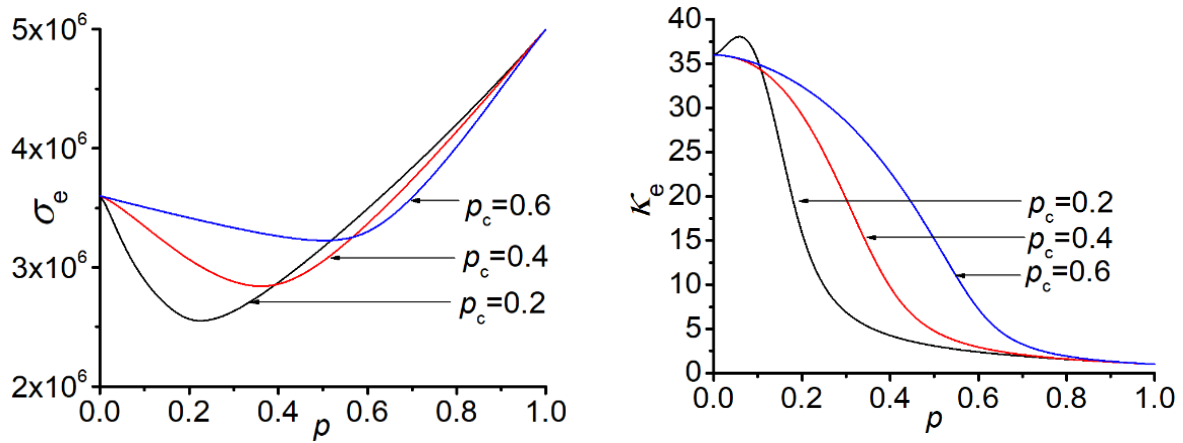

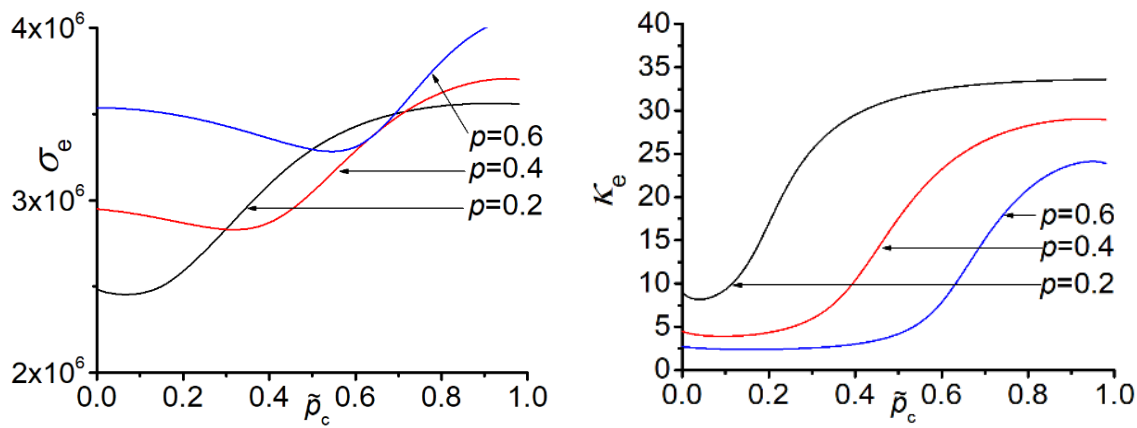

*Fig. 5. Dependences of effective thermoEMF and figure of merit on the percolation threshold $\tilde{p}_c$ at a given concentration $p$ of the first phase.*

The values of coefficients in the phases were selected the same in Fig.1.

**Conclusions**

This paper considers a modification of effective medium theory (self-consistency method) for thermoelectric phenomena in the case of a preassigned percolation threshold. Such a modification was first proposed in [11] for the description of galvanomagnetic phenomena and was used for the description of a series of experimental results [9-11,17]. In [18], the modification was used in the proposed approach of "mobile percolation threshold" to describe the magnetodielectric effect and the peculiarities of magnetic permeability of magnetoelastomers, in [19,20], to describe a gigantic magnetoelastic effect.

A similar modification is proposed here for a system of equations describing thermoelectric phenomena in macroscopically inhomogeneous two-phase composites.

The results obtained can be used for the description of thermoelectric properties of composites with different structures corresponding to different percolation thresholds.